\documentstyle[epsf,aps,12pt,floats,graphicx]{revtex}
\begin{document}
\draft
\tighten

\title{
Bond-disordered Anderson model on a two dimensional square lattice ---
chiral symmetry and restoration of one-parameter scaling
}

\author{Viktor Z.\ Cerovski}

\address{
Department of Physics and Astronomy, Michigan State University,
East Lansing, MI 48824
}

\date{Submitted to PRB on January 27, 2000.}

\maketitle

\begin{abstract}

Bond-disordered Anderson model in two dimensions on a square lattice
is studied numerically near the band center by calculating 
density of states (DoS), multifractal properties of 
eigenstates and the localization length.  
DoS divergence at the band center is studied and compared with Gade's result 
[Nucl.\ Phys.\ B {\bf 398}, 499 (1993)]
and the powerlaw.  Although Gade's form describes accurately DoS of finite size 
systems near the band-center, it fails to describe the calculated part of DoS 
of the infinite system, and a new expression is proposed.
Study of the level spacing distributions reveals that the state closest to 
the band center and the next one have different level spacing distribution 
than the pairs of states away from the band center.
Multifractal properties of finite systems furthermore show that scaling of 
eigenstates changes discontinuously near the band center.
This unusual behavior suggests the existence of a new divergent length scale,
whose existence is explained as the finite size manifestation of the
band center critical point of the infinite system, and the critical
exponent of the correlation length is calculated by a finite
size scaling.
Furthermore, study of scaling of Lyapunov exponents of transfer matrices of 
long stripes indicates that for a long stripe of any width there is an energy 
region around band center within which the Lyapunov exponents cannot be 
described by one-parameter scaling.  This region, however, vanishes in the
limit of the infinite square lattice when one-parameter scaling is 
restored, and the scaling exponent calculated, in agreement with the
result of the finite size scaling analysis.

\end{abstract} 


\section{Introduction}

A quantum particle moving in a random potential undergoes Anderson 
localization quantum phase transition in three dimensions with increasing of 
the strength of disorder \cite{Anderson58,Thouless74,Lee85,Kramer93}.
The order parameter characterizing the localized phase is the inverse 
localization length $\xi^{-1}$ \cite{Anderson80}, describing
the exponential decay of the envelope of eigenstates.  
When the critical point is being approached from the localized phase, 
localization length, which depends for a given energy only on the strength of
the disorder, increases with decreasing of the disorder strength 
and finally diverges as a power-law at a particular disorder strength.
Further decrease of disorder strength then makes the eigenstate extended 
throughout the whole system.
Simultaneously, on length scales smaller than the localization 
length, eigenstates exhibit multifractal scaling behavior characterized
by anomalous scaling of the inverse participation numbers
(for definitions and references, see Sec.\ \ref{MULTIFRACTALITY}).

This basic phenomena, together with the work of Thouless 
on the scaling of conductance in finite-size systems \cite{Licciardello75}, 
led to the scaling theory of localization \cite{Abrahams79,MacKinnon81}, 
one of the main consequences of which is
the absence of extended states in two dimensional disordered systems, 
with two dimensions being the lower critical dimension of the transition.
If the spin-orbit interaction is present, however, 
picture changes and two dimensional systems from symplectic ensemble 
exhibit localization transition even in two dimensions as opposed to systems
from orthogonal ensemble which have all states localized \cite{Lee85}.  
Presence of the strong magnetic field in two dimensional disordered systems, 
on the other hand, leads to a completely different behavior -- 
the integer quantum Hall effect (IQHE)  -- 
where critical states are present at the middle of each of 
disorder-broadened Landau levels \cite{Huckestein95}. 

Another class of models exhibiting localization properties different
from the systems mentioned above are systems with chiral
(particle-hole) symmetry.
Such systems are defined on a bipartite lattice with only hopping 
(off-diagonal or bond-) disorder.  Wegner was first
to realize the importance of this symmetry in disordered systems 
\cite{Wegner79,Gade93,Inui94}, and even one-dimensional 
systems with this symmetry are known to have peculiar properties, 
such as diverging DoS at the band center \cite{Theodorou76}, 
where the eigenstate decays as $\exp(-\gamma \sqrt{r})$ 
\cite{Fleishman77,Soukoulis81}, in contrast to one-dimensional 
site-disordered systems which have DoS bounded \cite{Wegner81} and all 
states localized. 

There are several models with chiral symmetry that have been extensively 
studied.  The simplest two, in the sense that only one orbital per site 
and nearest neighbor hopping are included, time-reversal symmetry present 
and the spin not relevant, are the Anderson bond-disordered model (ABD)
\cite{Theodorou76,Fleishman77,Soukoulis81,Soukoulis82,Eilmes98}
and the random Dirac fermion model (RDF) \cite{Hatsugai97,Morita97}. 
The main difference between these
two models is that, in the non-disordered case, ABD model has a line of points 
as the Fermi surface at half-filling while RDF has a point Fermi surface and 
linear dispersion of energies.  

This work is concerned with the ABD model on a square lattice of size $L$ and
periodic boundary conditions, defined by the Hamiltonian:
\begin{equation} \label{ABD}
   H = - \epsilon_0 \sum_{\langle i, j\rangle} 
         \left( t_{i,j} c_i^\dagger c_j + \mbox{H.c} \right),
\end{equation}
where brackets denote neighboring sites on the lattice,
$c_i$ is annihilation operator of the electron at site $i$,
and $t$'s are uniformly distributed random variables 
$t_{i,j}\in (1-2w,1)$, with $0 < w \le 1$.  
They represent random hopping energies
between nearest neighbors, expressed in units of energy $\epsilon_0$,
which is set to $1$ hereafter.

Interest in this model mainly comes from its unusual scaling properties
at the band center, where Soukoulis {\it et.\ al} \cite{Soukoulis82} have 
found critical state at the band center using Green's 
function\cite{MacKinnon81} 
and transfer matrix method (TMM) \cite{Pichard81}.
More recent TMM calculation by Eilmes {\it et.\ al}\cite{Eilmes98} 
confirmed this result with a higher accuracy and showed the validity of 
one-parameter scaling not too close to the band center.  Nevertheless, 
Miler and Wang \cite{Miler96} have found in their study of two models with 
chiral symmetry an apparent band of extended states near the band center,
and it remained unclear what is the fate of these states in the infinite
2D system.
Another study \cite{Cerovski99} yet showed that scaling exponent of 
the average participation number changed discontinuously near 
the band center, and the explicit dependence of this 
energy on the system size proposed by authors implied the existence 
of another diverging length scale in the problem. 
The last effect is rather subtle to calculate and led to a different
participation number scaling exponent of the ABD model at the band center
in Ref.~\cite{Eilmes98} compared to the one calculated here, 
as discussed in detail in Sec.\ \ref{MULTIFRACTALITY} below.
Furthermore, Brouwer {\it et al.} \cite{Brouwer99} have calculated 
conductance distribution of quantum wires described by (\ref{ABD}), 
and showed its non-universality and necessity to introduce an 
additional microscopic parameter.

It is thus goal of this paper to present a detailed study of 
the scaling of localization length on the approach to the band center
for an infinite 2D square lattice, and test the validity of 
one-parameter scaling, as well as to calculate
the multifractal properties of the electron probability density on 
length scales smaller than the localization length. 
Also a new analytical expression for the DoS of the infinite two-dimensional 
system near the band center is proposed.
This paper is organized as follows: Some general properties and exact results 
are presented in Sec.\ \ref{GENERAL}; Calculation of DoS is presented and 
analyzed in Sec.\ \ref{DOS}; Sec.\ \ref{SPACINGS} contains analysis of 
level spacing distributions between the nearest neighbors; 
Multifractal properties of eigenstates are studied 
in Sec.\ \ref{MULTIFRACTALITY}; 
Scaling of the Lyapunov exponents of transfer matrices of long strips and 
the scaling of localization length are studied 
in Sec.\ \ref{LOCALIZATION-LENGTH}; 
and, finally, Sec.\ \ref{CONCLUSIONS} summarizes results of this work.

\section{Some general properties of lattice Hamiltonians with chiral symmetry}
\label{GENERAL}

Suppose that the lattice is composed of two sublattices $A$ and $B$ with,
respectively, $N_A$ and $N_B$ sites.  The corresponding bond-disordered
Hamiltonian with chiral symmetry then has the form:
\begin{equation} \label{bipartite}
   H = \sum_{i\in A , j\in B} 
       \left( t_{i,j} c_i^\dagger c_j + \mbox{H.c} \right),
\end{equation}
It is easy to show that for every eigenstate $|\psi\rangle$  with energy 
$E$ there is an eigenstate with energy $-E$ with a wavefunction that
has the opposite sign at each site of one of the two sublattices.

If the total number of cites $N=N_A+N_B$ is odd and open boundaries condition 
is applied (in order to keep the symmetry), then, since all eigenstates
come in the opposite energy pairs, there will be exactly one state with
eigenenergy $0$.  This can be furthermore generalized, and if $m=N_A-N_B>0$,
there exist exactly $m$ zero-energy eigenstates that have vanishing amplitude
on the sublattice $B$ \cite{Inui94,Inuifootnote}.

On the other hand, if $m=0$, the electron has equal probability of occupying 
each of the two sublattices.
To show this, (\ref{bipartite}) is represented in the basis where 
the first and second half of basis vectors are eigenstates of 
the position operator on sites of sublattice $A$ and $B$, respectively. 
The Hamiltonian is then represented as
\begin{equation} \label{general}
   H = \left( \begin{array}{cc}  0 & M \\ M^\dagger & 0 \end{array}\right), 
\end{equation}
where $M$ is a square matrix of hopping elements from one sublattice 
to the other.  Eigenstate 
$|\psi\rangle = \left(\begin{array}{c} |\psi_A\rangle \\ |\psi_B\rangle 
\end{array}\right)$ satisfies:
\begin{equation} \label{A}
   E = \langle\psi | H | \psi\rangle \,=\, 
       2\,\mbox{Re}\,\langle\psi_A|M|\psi_B\rangle.
\end{equation}
On the other hand,
\begin{equation} \label{B}
   H|\psi\rangle = 
    \left(
      \begin{array}{c} M|\psi_B\rangle \\ M^\dagger|\psi_A \rangle \end{array}
    \right) =
    \left(
      \begin{array}{c} E|\psi_A\rangle \\ E|\psi_B\rangle \end{array}
    \right).
\end{equation}
From (\ref{A}) and (\ref{B}) now follows that, for ABD model, 
$\langle\psi_A|\psi_A\rangle = 1/2$.

In this work only even $L$ finite size systems on a square lattice with
periodic boundary conditions are studied, because one of the main goals of 
this work is to understand the vicinity of the critical point of ABD model on 
the infinite square lattice, which in turn has $m=0$, while the 
limit $L\to\infty$ for odd $L$ and open boundary conditions has $m=1$.

\section{Density of states near the band center} 
\label{DOS}

Density of states (DoS) is calculated by exact numerical diagonalization of 
finite size Hamiltonians for various $L$ for many configurations
of disorder, and binning of eigenenergies.  Obtained DoS for each system size, 
$\rho_L(E)$, are normalized to $1$.  
The $L$ dependent parts of such obtained $\rho_L(E)$ are then removed 
leaving $L$ independent DoS $\rho(E)$, which is therefore expected to be 
correct in the $L\to\infty$ limit.
The removal of finite size dependency is based on an observation that
DoS converges quickly away from the band center with increasing of $L$.  
Thus, only a small number of eigenenergies (up to 20) 
closest to the band center and corresponding DoS histograms has been calculated 
for each $L$.  The calculated $\rho_L(E)$ plotted on a single graph revealed 
that three bins closest to the band center are 
where the system size dependence sets in.  
Their removal thus led to the DoS $\rho(E)$ of the infinite system.

\begin{figure}[tb]
\begin{center}
\includegraphics*[width=6.1in]{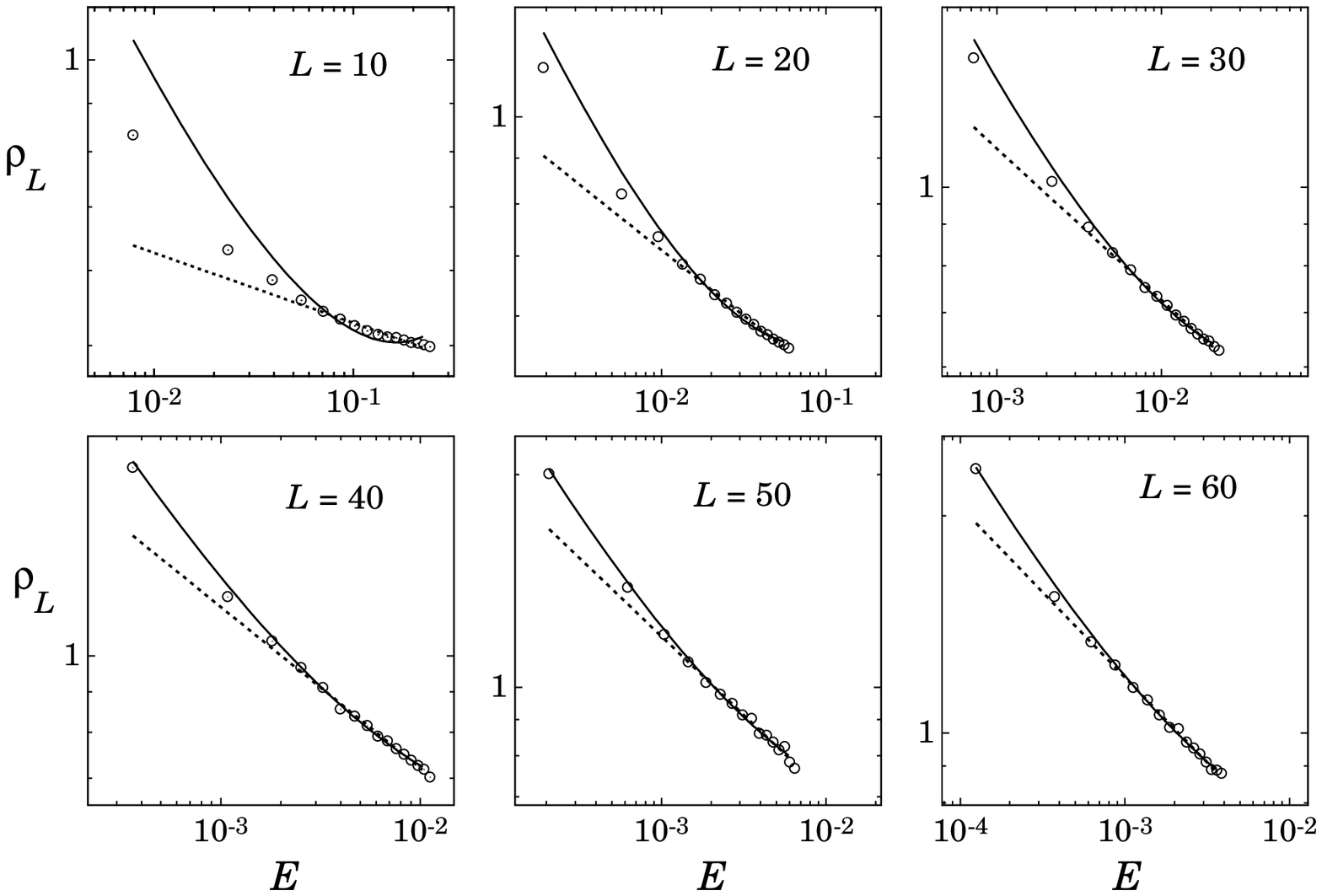}
\end{center}
\caption{
Density of states $\rho_L(E)$ of the ABD model for $w=1$ near the 
band center for system sizes $L=10,...,60$ in log-log plot.  
Full lines are fits to the Gade's form (\ref{Gade}), while dotted lines are
fits to the powerlaw.  Fit to the Gade's form is done for all the points except
the one closest to the band center, while fit to the powerlaw is done for all 
the bins except the three bins closest to the band center, which are the 
$L$-dependent parts of $\rho_L(E)$.
}\label{DOS-fits}
\end{figure}

Results for $\rho_L(|E|)$ are given in Fig.~\ref{DOS-fits}, 
for system sizes $L=10,20,...,60$ and number of disordered configurations
ranging, respectively, from $160000$ to $4100$.  
They are fitted to a powerlaw divergence $\rho_L(E) = C_L |E|^{-\alpha_L}$, 
as well as to the Gade's result \cite{Gade93}:
\begin{equation} \label{Gade}
   \rho_L(E) \,=\, C_L {1\over|E|}\exp\left( -\kappa_L\sqrt{-\ln |E|} \right).
\end{equation}
All the calculations were done for several different number of bins, and 
the obtained values of fitting parameters were the same within error bars.

\begin{figure}[tb]
\begin{center}
\includegraphics*[width=6.1in]{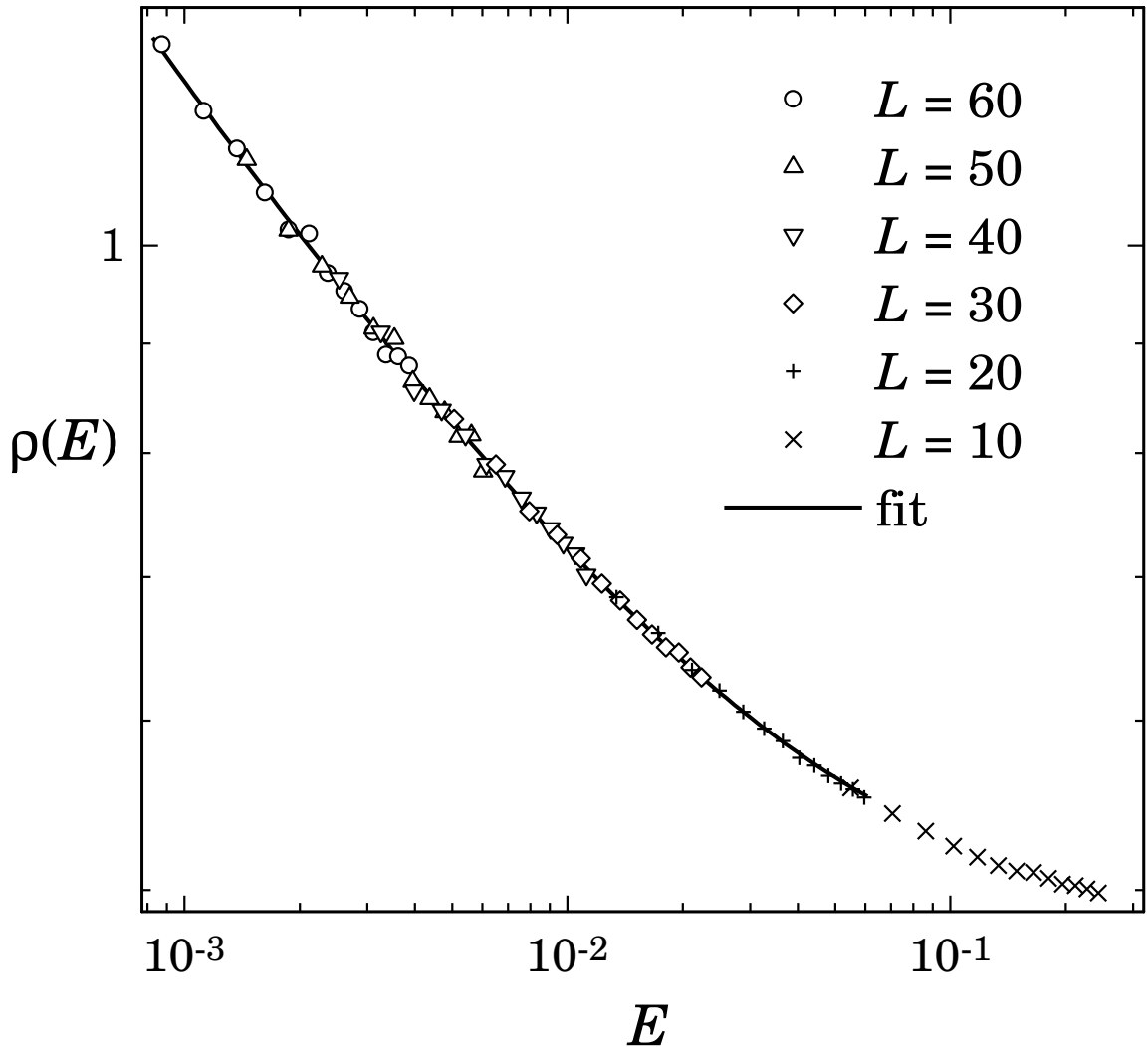}
\end{center}
\caption{Density of states of the ABD model for $w=1$ near the band center.
The graph is obtained from the data in Fig.~\ref{DOS-fits} by removing the 
three bins closest to the band center, leaving the $L$-independent $\rho(E)$.
The fit is $\rho(E)=C\exp\left(-\kappa\sqrt{-\ln |E|}\right)/\sqrt{|E|}$,
with $\kappa=1.345\pm 0.005$ and $C=1.30\pm 0.03$.
} \label{DoS}
\end{figure}

The figure \ref{DOS-fits} shows that (\ref{Gade}) describes $\rho_L$ 
very accurately for $L\ge 40$, including the size-dependent part. 
The powerlaw, on the other hand, also describes the data accurately for
same system sizes, but fails to describe the $L$-dependent part of $\rho_L$.
Despite this, neither of the two forms describe the whole $\rho(E)$ 
accurately.  Instead, the expression found to best fit the obtained 
$L$-independent DoS, given in Fig.~\ref{DoS},  is 
\begin{equation}\label{new-DoS}
  \rho(E)\,=\,C {1\over {\sqrt |E|}}\,\exp\left(-\kappa\sqrt{-\ln |E|}\right),
\end{equation}
with $\kappa=1.345\pm 0.005$ and $C=1.30\pm 0.03$, represented by the full 
line in the same figure.  The observed range in which (\ref{new-DoS}) 
is accurate is for all the energies studied smaller than $6\times 10^{-2}$.

\section{Distribution of the nearest-neighbor level spacings} \label{SPACINGS}

In the localized regime, an eigenstate is determined mainly by 
a local configuration of disorder where the wavefunction is localized, 
and two eigenstates close in energy are spatially far apart.  
Level repulsion is therefore absent
and the distribution of the nearest neighbor level spacings 
$s\equiv E_{i+1}-E_i$ is Poissonian \cite{Shklovskii93},
\begin{equation} \label{Poisson}
  D_P(s)\,=\, {1\over\delta} \exp\left(-{s\over\delta}\right).
\end{equation}
where $\delta \equiv \langle s\rangle$ is the mean level spacing.

In the delocalized phase, on the other hand, eigenstates are extended 
throughout the system and level repulsion becomes significant for eigenstates
with close energies.  In the infinite 3D Anderson site-disordered model (ASD),
\begin{equation}\label{ASD}
   H\,=\,\sum_i \epsilon_i \, n_i \,-\, 
         \sum_{\langle i,j\rangle}
             \left( c_i^\dagger c_j + c_j^\dagger c_i\right),
\end{equation}
with uniformly distributed $\epsilon_i \in (-W/2,W/2)$, 
distribution of level spacings becomes
that of Gaussian orthogonal ensemble (GOE) \cite{Shklovskii93},
very accurately described by the Wigner surmise
\begin{equation} \label{Wigner}
   D_W(s)\,=\,{\pi\over 2}{s\over\delta^2}
       \exp\left( -{\pi\over 4}\left({s\over\delta}\right)^2 \right).
\end{equation}

In finite-size systems, localized states are on average at a distance $L$ 
rather than infinitely far apart.  This leads to a repulsion between adjacent 
energy levels and non-universal distribution $D_L(s)$ .
Shklovskii~{\it et.~al.} \cite{Shklovskii93} have shown that $D_L(s)$
of the 3D site-disordered Anderson model exhibits linear dependence on $s$
characteristic for $D_W(s)$ for small $s$ and exponential tail characteristic
for $D_P(s)$ for large $s$.  They were able, from the finite size
scaling analysis of the tail, to accurately determine the critical point and
exponent.  In the infinite size limit, they have recovered not only $D_P(s)$ 
in the insulating phase and $D_W(s)$ in the conducting phase, 
but also a system size independent
non-universal distribution at the critical point
which was furthermore shown by Braun {\it et.~al.}\cite{Braun98} to be 
dependent on boundary conditions.  
This method was also used for an accurate determination 
of localization length in two dimensional ASD model \cite{Zharekeshev96},
confirming the absence of delocalized states following the scenario of 
the insulating phase from Ref.~\cite{Shklovskii93} described above.

To see the effect of symmetry of the Hamiltonian (\ref{bipartite}) on 
the distribution of level spacings, let us for a moment consider $i$-th 
eigenenergy $E_i$ of the ASD model.  Upon averaging over disorder, the
$E_i$ will be distributed between $E_i^{min}$ and $E_i^{max}$
according to some distribution.  Some of the eigenenergies, for $i$
close to $N/2$, will have $E_i^{min} < 0 < E_i^{max}$.
This is, however, forbidden for eigenstates of the ABD model
since every $E_i$ of (\ref{bipartite}) is negative for $i<N/2$ 
and positive for $i>N/2$.  This means that eigenenergies of (\ref{bipartite}) 
close to the band center are effectively pushed away from it due to 
the symmetry.  
If $L$ is much smaller than the localization length, states will be 
repelled among themselves due to their large spatial overlap. 
But the two states closest to the band center, being simply related to each 
another by the symmetry, will not repel at all, {\it i.e.} the state 
closest to the band center is at the (high energy) end of the spectrum.
Thus, these two states are distributed around zero, where distributions 
of all other individual levels go to zero.

\begin{figure}[tb]
\begin{center}
\includegraphics*[width=6.1in]{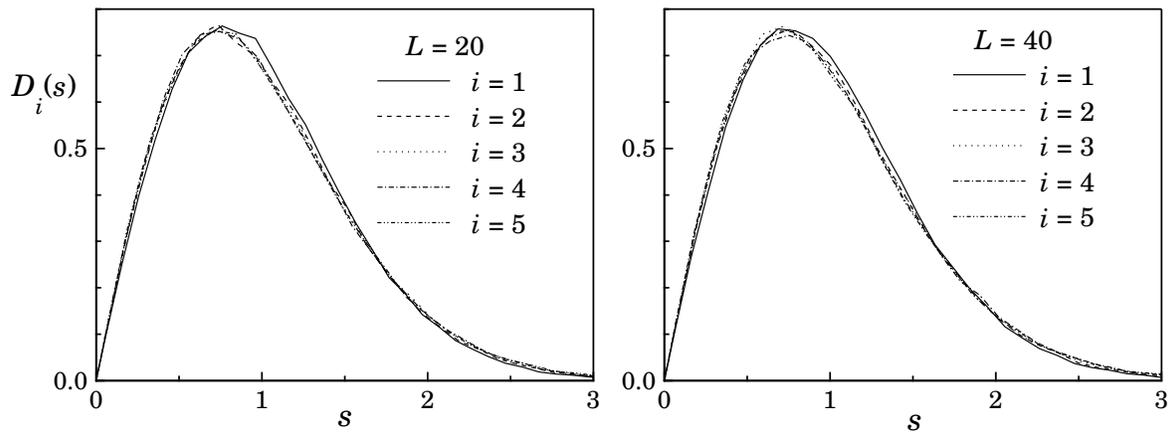}
\end{center}
\caption{Distributions of level spacings (after unfolding of the spectrum)
$D_i(s)$ between $i$ and $i+1$ level, counted from the band center.  
$D_1(s)$ is distinctly different than other distributions due to the
presence of the symmetry.
} \label{NN}
\end{figure}

To explore consequences of this simple analysis, the level spacing 
distribution is calculated between each pair of adjacent levels separately.
Let us denote by $D_i(s)$ level spacing distribution between $i$ and $i+1$ 
energy level after unfolding of spectrum \cite{Haake91},
{\it i.e.} expressing level energies in units of mean level spacing,
where $0<E_1<E_2<...<E_{N/2}$.  Figure~\ref{NN} shows $D_1(s),...,D_5(s)$, 
for $L=20,40$ and, respectively $150000,120000$ configurations, 
and it can be seen that $D_1(s)$ is distinctly different than 
$D_2(s), ..., D_5(s)$.  
The same effect was also present for $L=10,30$, while for $L=50$ and $60$ 
the number of disorder configurations was insufficient
for accurate enough determination of individual $D_i(s)$ \cite{preliminary}.
This illustrates how the presence of chiral symmetry can profoundly influence
spectral characteristics near the band center, despite the fact that
DoS of ABD and ASD models seem to have same shape for adequately
chosen pairs of disorder parameters $w$ and $W$ away from the band center
(and after rescaling of $\epsilon_0$) \cite{Eilmes98}.

\section{Multifractality of eigenstates} \label{MULTIFRACTALITY}

Eigenstate of an electron in the random potential fluctuates from site to site 
and it was proposed that the eigenstate at the mobility edge in disordered 
systems in general should have fractal structure\cite{Aoki83/86}, and shown 
that even localized states in one and two dimensions 
exhibit fractal character on length scales smaller than the localization length 
\cite{Soukoulis84,Fal'ko95}.

Inverse participation numbers $Z_q$ (IPN) are particularly convenient 
quantities to describe scaling properties of probability distribution 
of the electron. IPN of an eigenstate $\Psi$ are defined as:
\begin{equation} \label{IPN-psi}
   Z_q(\Psi) \equiv \sum_{i=1}^{L\times L} |\Psi({\bf r}_i)|^{2q}.
\end{equation}

Intuitively, their meaning can be seen by looking at the participation number 
$Z_2(\Psi)^{-1}$ :  it is equal to 1 for a state localized at one
site and to $N$ for plane-waves.  Participation number thus gives generally
the number of sites at which the wavefunction is significantly 
different than zero.
Participation numbers $Z_q(\Psi)^{-1}$ then generalize this by 
giving the number of sites where probability distribution of electron is 
very high (for large positive $q$'s), very low (for large negative $q$'s), 
and in between these extrema, continuously parameterized by $q$.

More convenient, with the advantage of being defined as averages over disorder 
at a given energy $E$, are IPN defined as functions of $E$ and system size $L$,
\begin{equation} \label{IPN}
   Z_q(L,E)\equiv\left\langle\,Z_q(\Psi)\,\delta(E(\Psi)-E)\,\right\rangle,
\end{equation}
where the brackets denote averaging over disorder.
$Z_q(E,L)$ can be numerically calculated by averaging (\ref{IPN-psi}) over all
eigenstates from $M$ configurations of disorder belonging to an energy 
interval of width $\Delta E$ around $E$, and studying the limit $\Delta E\to 0$
for large $M$ \cite{Cerovski99}.

Wegner \cite{Wegner80} pioneered this kind of investigations, and 
Castellani and Peliti \cite{Castellani86} proposed that eigenstates near the
critical point are multifractal on length scales smaller than $\xi$.
The most important feature of IPN of eigenstates is their scaling with
system size and energy \cite{Wegner80,Castellani86}:
\begin{eqnarray}\label{ZqScaling}
   Z_q(L,E) & \sim & L^{-\tau_q},     \label{tau_q} \\
   Z_q(L,E) & \sim & |E-E_c|^{\pi_q}, \label{pi_q}
\end{eqnarray}
where $E_c$ is the critical energy.  The former scaling is present at any $E$
for $L\ll\xi(E)$, while the latter holds in the critical region of the
transition \cite{logarithmic}.

Within the framework of multifractality \cite{Hasley86,Jensen87}, 
electron probability density is characterized by several quantities 
that can be derived from
$\tau_q$ --- the generalized dimension $D_q$ and the singularity strength 
$\alpha_q$ of the $q$-th singularity with the fractal dimension $f_q$ :
\begin{equation}\label{multifractal-definitions}
   (q-1)D_q      \equiv  \tau_q                   \;,\;
   \alpha_q      \equiv  {d \tau_q \over d q}     \;,\;
   f_q(\alpha_q) \equiv  \alpha_q\,q\,-\,\tau_q.
\end{equation}
$D_q$ represents generalization of the fractal dimension,
and it is constant and equal to the fractal dimension for ordinary fractals, 
while $f_q(\alpha_q)$ is the singularity strength spectrum describing 
multifractal as an interlaced set of fractals with fractal dimensions
$f_q$, where the measure on the $q$-th fractal scales 
as a powerlaw with exponent $\alpha_q$.
These quantities have several general properties: 
$D_0$ is the fractal dimension of the support (two in this work); $D_1$
is called information dimension since it describes scaling of the entropy of 
the measure\cite{D_1 remark}, 
and there exist finite $D_{min}=D_{q\to\infty}$ and $D_{max}=D_{q\to-\infty}$.

This work is concerned mainly with the spectrum of generalized dimensions
$D_q$ characterizing the spatial structure of eigenstates on the length scales
smaller than the localization length, while the properties of $\pi_q$ will be
discussed elsewhere.  Before detailed discussion of the results, 
an overview of the main results of this part of the paper is given.  
Scaling of IPN at the band center is calculated first, 
and shown that scaling properties (that is, whole spectrum of generalized 
dimensions $D_q$) 
changes discontinuously near the band center, at an energy
$E'(L)$ for a range of $L$ studied.  
It is then shown that $E'$ can be quite accurately identified with the half 
of the width of the energy range around band center
within which two states occur on average in the ensemble of 
disordered systems. 
Existence of this energy reveals the existence of a length scale $\xi'(E)$, 
diverging when $E\to 0$, that is the system size at which IPN change 
their scaling properties from one powerlaw dependence on $L$ to another.  
This change is then explained as a finite size manifestation of the critical 
point, and the critical exponent calculated by a finite-size analysis.

\begin{figure}[tb]
\begin{center}
\includegraphics*[width=6.1in]{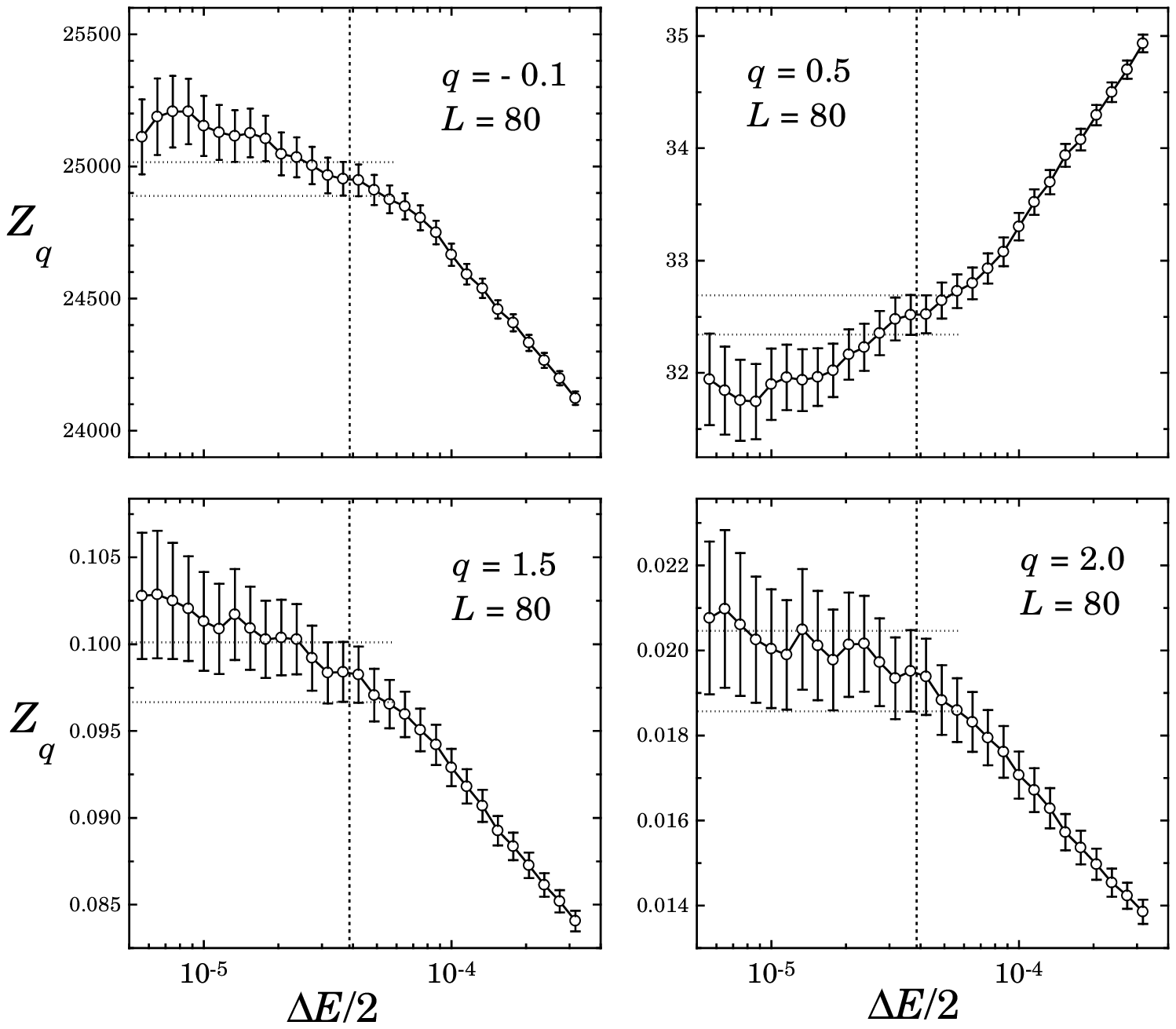}
\end{center}
\caption{Dependence of the average IPN for several different $q$'s and $L=80$
on the bin size $\Delta E$.  The vertical dashed lines represent energies 
$E_2(L)/2$ (see text for discussion).
} \label{convergence-q}
\end{figure}

Calculation of $Z_q(E,L)$ starts with calculation of $Z_q(E,L,\Delta E)$, 
which is just IPN averaged over all eigenstates from an energy interval  
$(E-\Delta E/2,E+\Delta E/2)$, taken from $N_\Omega$ realizations of disorder,
followed by studying the limit $\Delta E \to 0$ \cite{Cerovski99}. 
Results at the band center for system $L=80$ and several different $q$'s, 
are presented in Fig.~\ref{convergence-q}.  The error bars in the figure are 
taken to be the standard deviation of average value.

\begin{figure}[tb]
\begin{center}
\includegraphics*[width=6.1in]{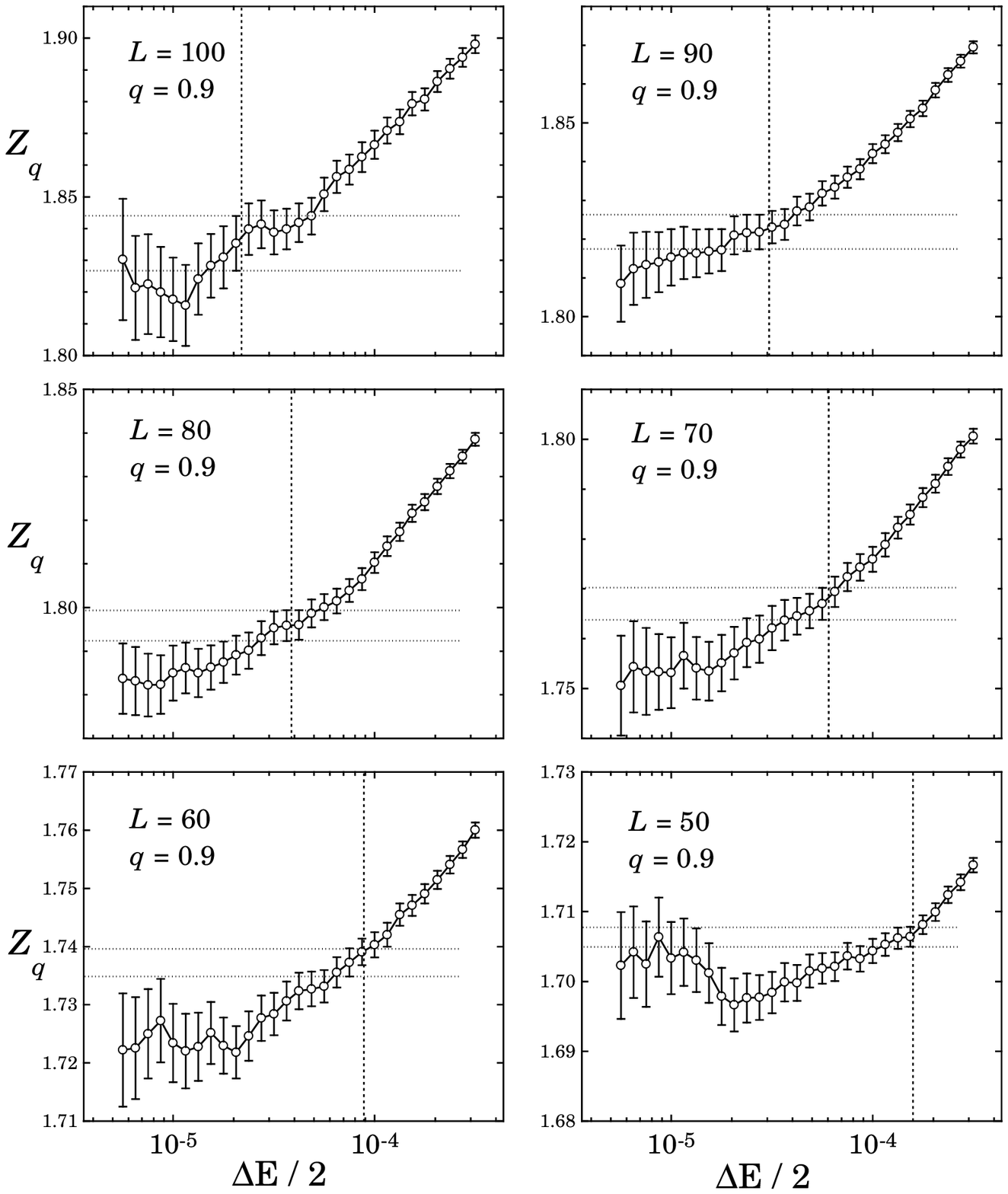}
\end{center}
\caption{Dependence of the average IPN for several different $L$ and $q=0.9$
on the bin size $\Delta E$. The vertical dashed lines represent energies 
$E_2(L)/2$ (see text for discussion).
} \label{convergence-L}
\end{figure}

The figure suggests the existence of an energy $E'$ independent of $q$ 
(and therefore defined by the whole multifractal measure) such that decreasing 
$\Delta E$ below $E'$ does not change $Z_q(E,L,\Delta E)$ significantly.
Decreasing of $Z_q$ to a smaller extend, however, is still present 
for $\Delta E < E'$, and the main source of this is the mismatch between 
average and typical value of IPN at a given energy.  
Thus, the effect should become smaller as the number of 
disorder configurations that are averaged over is increased, and
$Z_q(E,L,\Delta E<E') \approx Z_q(E,L)$ up to the corresponding 
statistical error.  
This can be seen in Fig.~\ref{convergence-q} and \ref{convergence-L},
where values of $Z_q$ for $\Delta E<E'$ are approximately constant
within the error bars (as indicated by the horizontal dashed lines),
while for $\Delta E>E'$ there is approximately
linear dependence of $Z_q$ on the bin size $\Delta E$.
In this sense Figure~\ref{convergence-L} suggests that $E'$ exists 
for all the systems studied (and can be shown to be independent of $q$ 
for each of them analogously as shown in Fig.~\ref{convergence-L}).

Analogous analysis is carried out for energies away from the band center 
and $Z_q(E,L)$ determined accordingly, where it turns out that the convergence 
for these energies is slightly easier to establish and occurs at larger 
$\Delta E$ than at the band center.  Such obtained energy intervals used in 
calculations of IPN for different $E$ were small enough so that there was 
practically no overlap among them.

The vertical dashed lines in Fig.~\ref{convergence-q} and 
Fig.~\ref{convergence-L} represent half of the energy $E_2(L)$, defined as 
the width of the energy interval around zero in which every system from 
the ensemble of disordered systems has two states on average\cite{Cerovski99},
\begin{equation} \label{divergent}
  1\,=\,L^2\,\int_{E_2/2}^0\rho_L(\epsilon)\,d\epsilon.
\end{equation}
From the figures one can see that $E'\approx E_2/2$ for all system sizes 
studied except $L=40$ (the smallest system studied, not shown in 
Fig.~\ref{convergence-L}), where the convergence seems to be somewhat slower.
Equation (\ref{divergent}) defines a new length scale $\xi'(E)$ that
can be described as the system size $L$ for a given energy $E$ such that 
the number of states within the energy interval between $-E$ and $E$ is 
two on average.  This can be defined as
\begin{equation}\label{xi'}
   \xi'(E) \,=\, E_2^{-1}(E),
\end{equation}
where $E_2^{-1}(E)$ is the inverse function of the $E_2(L)$, which is defined 
by (\ref{divergent}).   Since $E_2(L)$ goes to zero when
$L\to\infty$, the new length scale diverges when $E\to 0$.

It is tempting to integrate results for $\rho_L$ from Sec.~\ref{DOS} 
to obtain $\xi'(E)$ explicitly.
This, however, does not give the correct result since the whole analysis of
DoS from Sec.~\ref{DOS} is done for energies larger than the width of the
distribution of two states closest to the band center, 
which in turn defines $\xi'$ in (\ref{xi'}). 
In other words, there is an energy cutoff, vanishing when $L\to\infty$, 
below which the fits are not accurate, most obviously seen by noticing
that all of the assumed analytical forms of $\rho_L$ are diverging at 
the band center, while the actual $\rho_L$ is not.

Nontrivial feature connected with the existence of the new length scale $\xi'$
is that scaling exponents $\tau_q(E)$ are different for $L\lesssim \xi'(E)$ and 
$L\gtrsim\xi'(E)$.  
This is a generalization of findings from Ref.~\cite{Cerovski99},
where a deviation from the powerlaw scaling of the average participation number
for different $\Delta E$
at the band center appeared whenever $\Delta E$ exceeded $E'(L)$ in
several models with chiral symmetry.  
It can be straightforwardly shown that same happens with scaling of 
$Z_q(E=0,L,\Delta E)$.
This suggests that a new scaling characteristic should be attributed 
to the two states closest to the band center,  
and that $E'$ scales as $E_2$, with a coefficient of 
proportionality close to one.  This also implies that the corrections to 
the constant $Z_q(E,L,\Delta E)$ for $\Delta E < E'$ are small.
Therefore, the length at which change of scaling occurs should be close to
and depend on the energy proportionally to $\xi'(E)$, 
while the change of scaling should be a narrow crossover,
as opposed to much broader crossover from powerlaw to constant IPN that 
occurs at the termination of multifractal scaling (\ref{tau_q})
for $L \approx \xi$.

\begin{figure}[tb]
\begin{center}
\includegraphics*[width=6.1in]{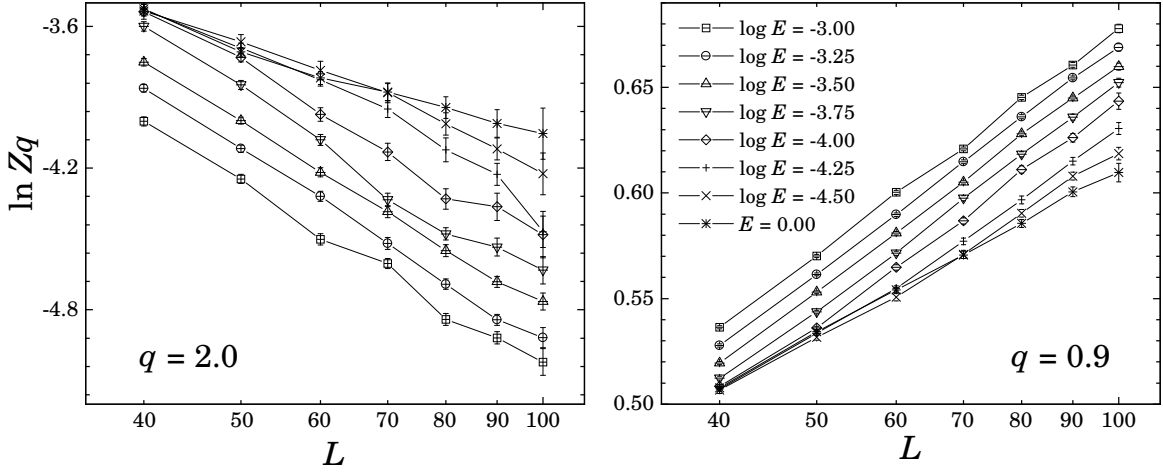}
\end{center}
\caption{Dependence of the average IPN on system size for several energies
near the band center for $q=0.9, 2$.
} \label{lnZqvslnL}
\end{figure}

This is compatible with the results presented in Fig.~\ref{lnZqvslnL}, 
which gives calculated $Z_q(L,E)$, for $q=0.9, 2$ as well as for 
all $L$ and $E$ studied.   
Smaller $q$'s allow for more accurate determination of the scaling 
properties, and it can be seen from the data that for both $q$'s there are
two characteristic scaling behaviors --- one occurs at the band center and
nearby energies for smaller $L$, while the other scaling holds
at energies further away from the band center, and for larger $L$ at energies 
close to the band center.  

\begin{figure}[tb]
\begin{center}
\includegraphics*[width=6.1in]{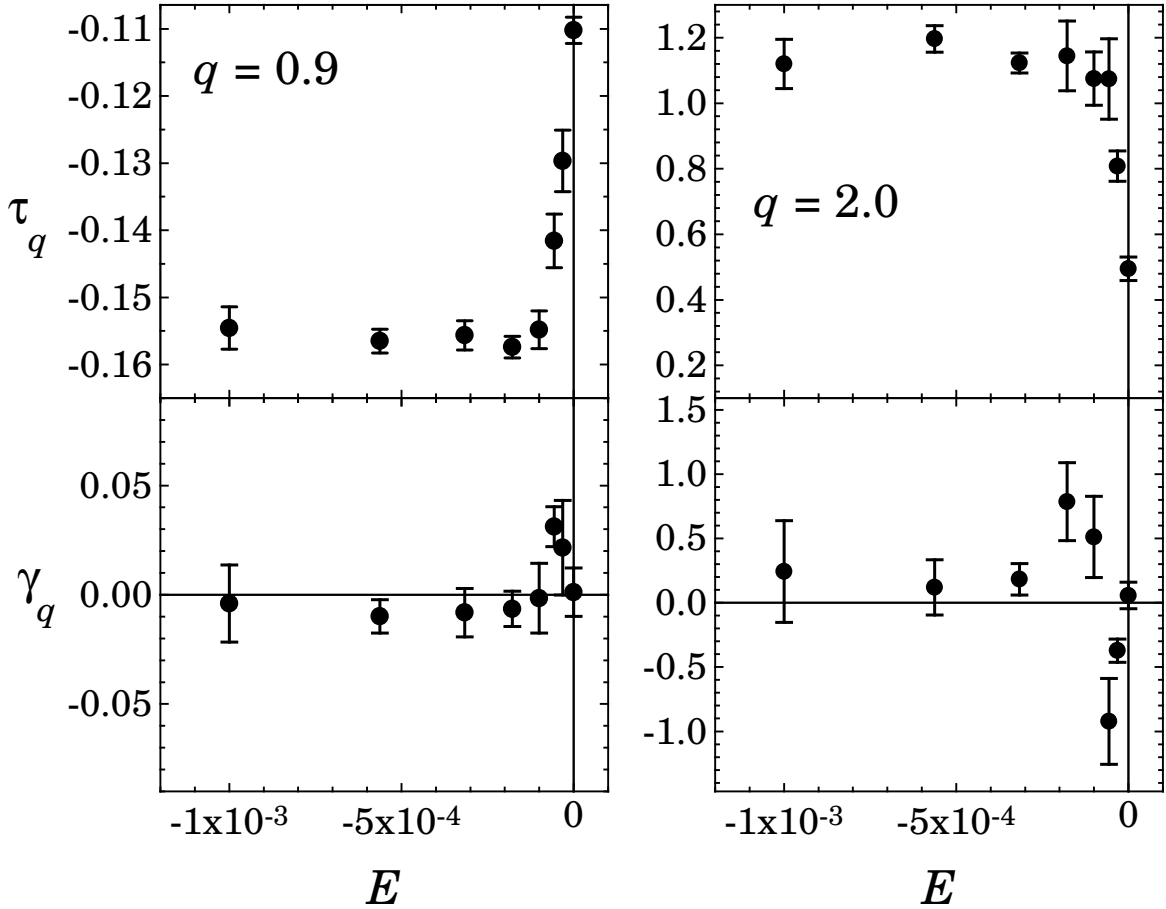}
\end{center}
\caption{Scaling exponent of IPN, $\tau_q$, and the goodness of
fit coefficient $\gamma_q$, calculated from the data
in Fig.~\ref{lnZqvslnL} (except $L=40$ results).  
} \label{tau-gamma}
\end{figure}

Scaling exponents $\tau_q$ are determined form the linear regression
of the data for system sizes $L=50,60,...,100$.  
Goodness of the powerlaw fit is quantitatively characterized by a
coefficient $\gamma_q(E)$ next to the quadratic term from 
an additional quadratic fit of the data. Such obtained 
$\tau_q$ and $\gamma_q$ from the data in Fig.~\ref{lnZqvslnL} are 
presented in Fig.~\ref{tau-gamma}.  
Results show the existence of three different cases: 
(i) away from the band center, $\gamma_q\approx 0$ and $\tau_q$ is independent 
of $E$, indicating powerlaw dependence of IPN on $L$; 
(ii) approaching the band center, $\gamma_q$ becomes different than zero, 
indicating that powerlaw is not obeyed.  
This is due to the emergence of the new scaling for system sizes 
$L \gtrsim \xi'(E)$ discussed above and present in Fig.~\ref{lnZqvslnL}; 
and (iii) for $E=0$ powerlaw is obeyed again ($\gamma_q\approx 0$), 
but with a different $\tau_q$ than for energies away from the band center.

\begin{figure}[tb]
\begin{center}
\includegraphics*[width=6.1in]{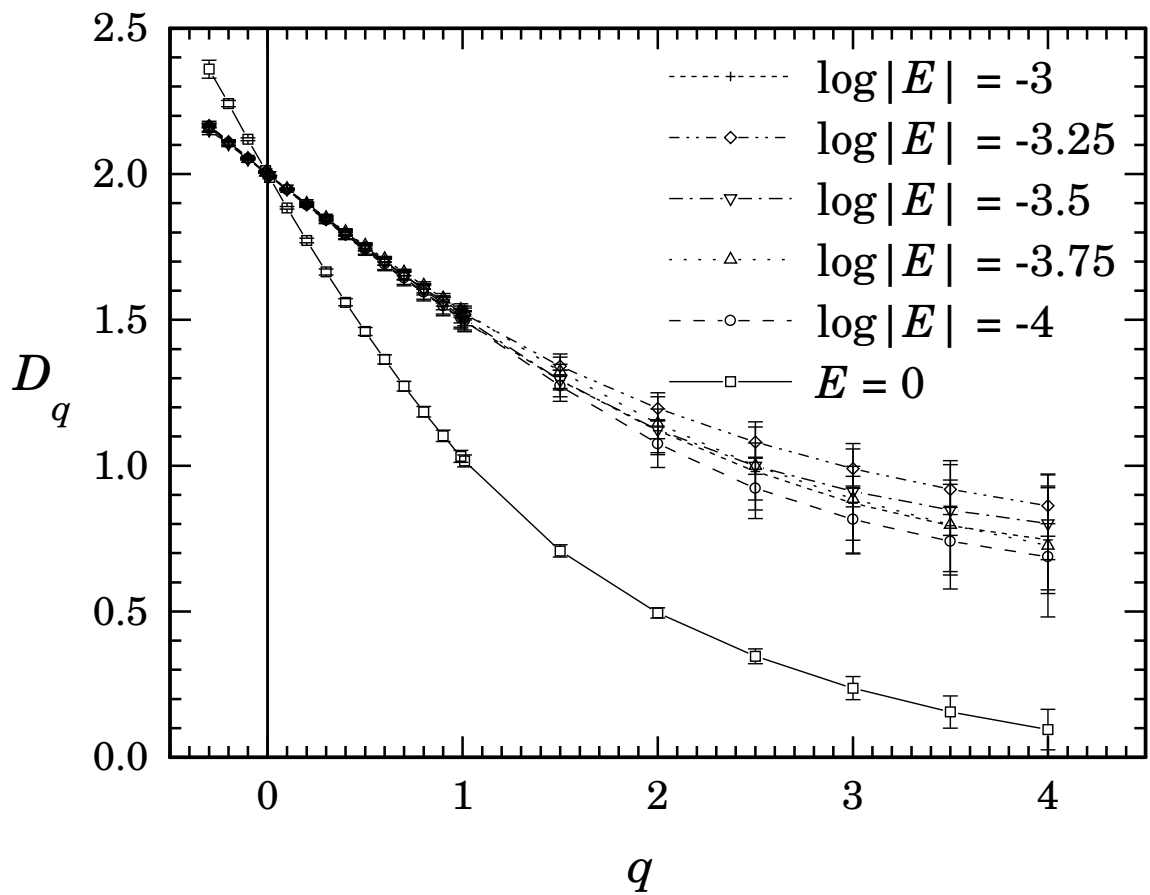}
\end{center}
\caption{Multifractal spectrums $D_q$ at the band center and nearby energies.
}\label{Dq}
\end{figure}

Discontinuity of $\tau_q(E)$ for all the other $q$'s studied naturally leads 
to two different spectra of generalized dimensions, and Fig.~\ref{Dq} shows
calculated $D_q$ for all energies except the three nonzero energies closest 
to the band center (which cannot give $\tau_q$ from the fitting procedure 
used here due to the change of scaling properties discussed above).  
In particular, the participation number 
grows with the number of sites $L^2$ as a powerlaw with exponents 
$\beta(E=0)=0.25\pm 0.02$, and $\beta(E\ne 0)= 0.55\pm 0.05$.  This should be
compared with the result for ABD model of Ref.~\cite{Eilmes98}, 
$\beta(E=0)=0.50\pm 0.06$.  It is easy to explain this discrepancy since the
bin size $\Delta E = 4\times 10^{-4}$ used in Ref.~\cite{Eilmes98} was, 
depending on the system size, roughly over an order of magnitude too large
to detect the correct scaling behavior, and therefore $\beta(E\ne 0)$ was 
obtained instead.

$D_q$ is calculated for only three negative values of $q$, mainly because IPN 
for negative $q$'s are determined mostly by the parts of eigenstates with 
the smallest probability to find the electron, which in turn acquire 
the highest relative error during numerical diagonalization of the Hamiltonian.
Difficulties in calculating $D_q$ in this regime even arose 
suspicion that multifractality might break down for negative $q$ 
\cite{Evangelou90}.  
It is thus important to show that $D_q$ is defined for negative $q$'s 
as well as for positive ones. 
Accuracy of all the calculations of IPN done in this section can be 
straightforwardly improved by increasing the number of configurations of 
disorder that was averaged over.  
This would lead to smaller error bars of all the
quantities calculated, as well as to the wider range of $q$'s for which
$D_q$ can be calculated.

Results of this section give the following picture of the scaling of IPN with 
system size: for any energy $E$ close enough to the band center,
there exist two powerlaw scalings: one for $L_0<L\lesssim\xi'(E)$ 
described by the set of exponents $\tau_q(E=0)$, 
and another one for $\xi'(E)\lesssim L\ll\xi(E)$, 
described by a {\em different} set of exponents $\tau_q(E\ne 0)$, which 
leads to the two different spectra of generalized dimensions $D_q$.

\begin{figure}[tb]
\begin{center}
\includegraphics*[width=6.1in]{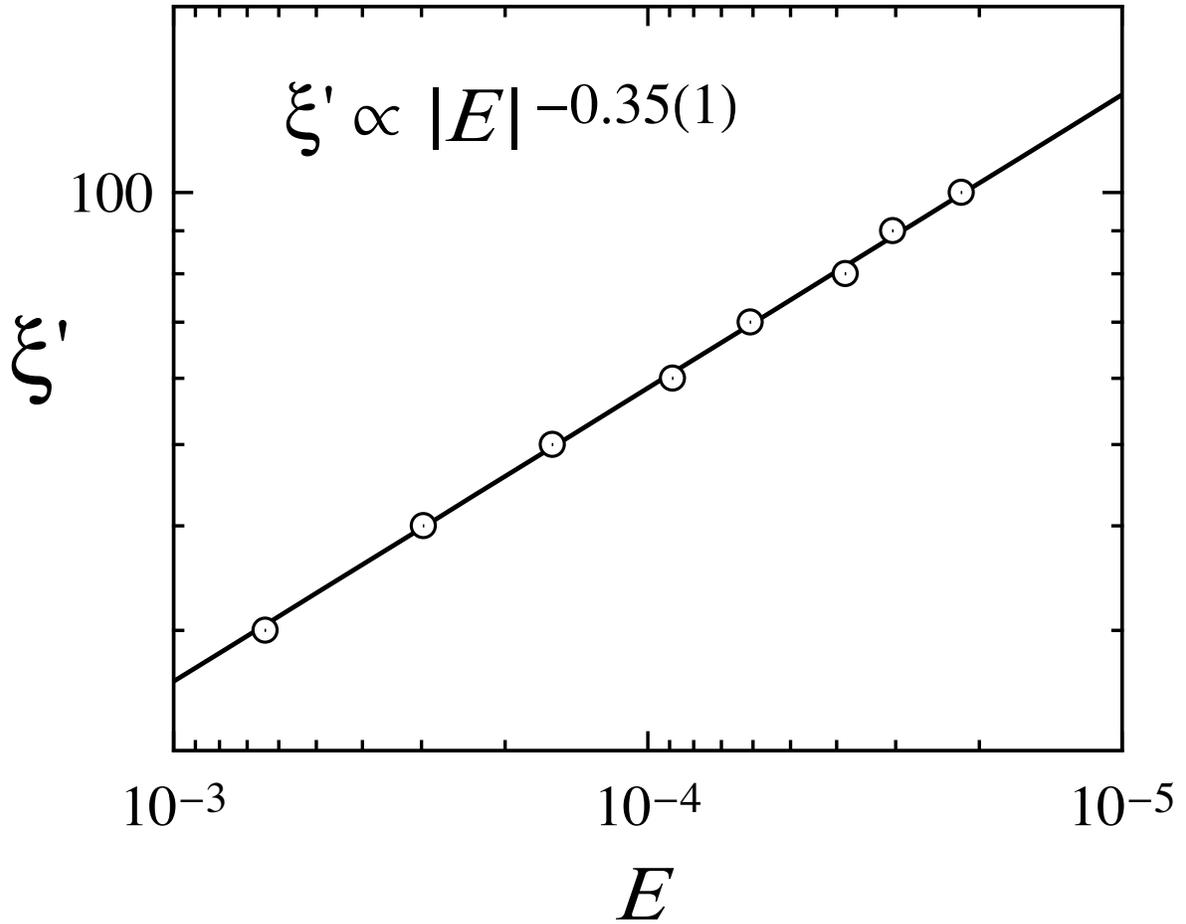}
\end{center}
\caption{Dependence of the new length scale on energy near the band center,
as determined from the Eq.\ (\ref{xi'}).
}\label{xi'(E)}
\end{figure}

Dependence of the new length scale on energy, $\xi'(E)$, can be easily 
determined from (\ref{xi'}) by integrating the actual numerical data for 
$\rho_L(E)$, and the result, obtained from Fig.~\ref{xi'(E)}, 
gives $\xi'(E)\propto |E|^{-\nu'}$, with
\begin{equation}\label{NU'}
   \nu' = 0.35\pm 0.01\; .
\end{equation}

The meaning of this new length scale and corresponding exponent $\nu'$ can be 
understood by assuming that the additional scaling of the two states is due to 
the finite size effect of ``smearing'' of the $E=0$ critical point of
the infinite system, because exactly the two states closest to the band center
become critical when $L\to\infty$.  
The critical energy then changes by $\Delta \epsilon$ 
which is, in a finite system of size $L$ (in units of the lattice spacing), 
equal to $\Delta E_2(L)\propto L^{-\delta}=L^{-1/\nu'}$.
If furthermore $\xi\propto |E|^{-\nu}$ in the infinite system,
the shift $\Delta \epsilon_c$ of the critical energy in the system of size 
$L$ is $\Delta \epsilon_c \propto L^{-1/\nu}$, from the general theory of the
finite size scaling \cite{Cardy96}.  
Therefore, $\nu'$ equals $\nu$, the critical exponent of the correlation length 
of the infinite two-dimensional system.
It should be noticed that the finite size scaling applied here is 
somewhat different than usual, where 
$\Delta \epsilon_c \equiv |E-E_c|/E_c$ \cite{Cardy96}.  Here,
$E_c=0$ and $\Delta \epsilon_c \equiv |E-E_c|$, where both energies are 
expressed in units of $\epsilon_0$, as discussed in the Introduction.
Thus, energy $\epsilon_0$ appears in 
the denominator of $\Delta\epsilon_c$ rather than the critical energy itself.

\section{Localization length near the band center}
\label{LOCALIZATION-LENGTH}

\begin{figure}[tb]
\begin{center}
\includegraphics*[width=6.1in]{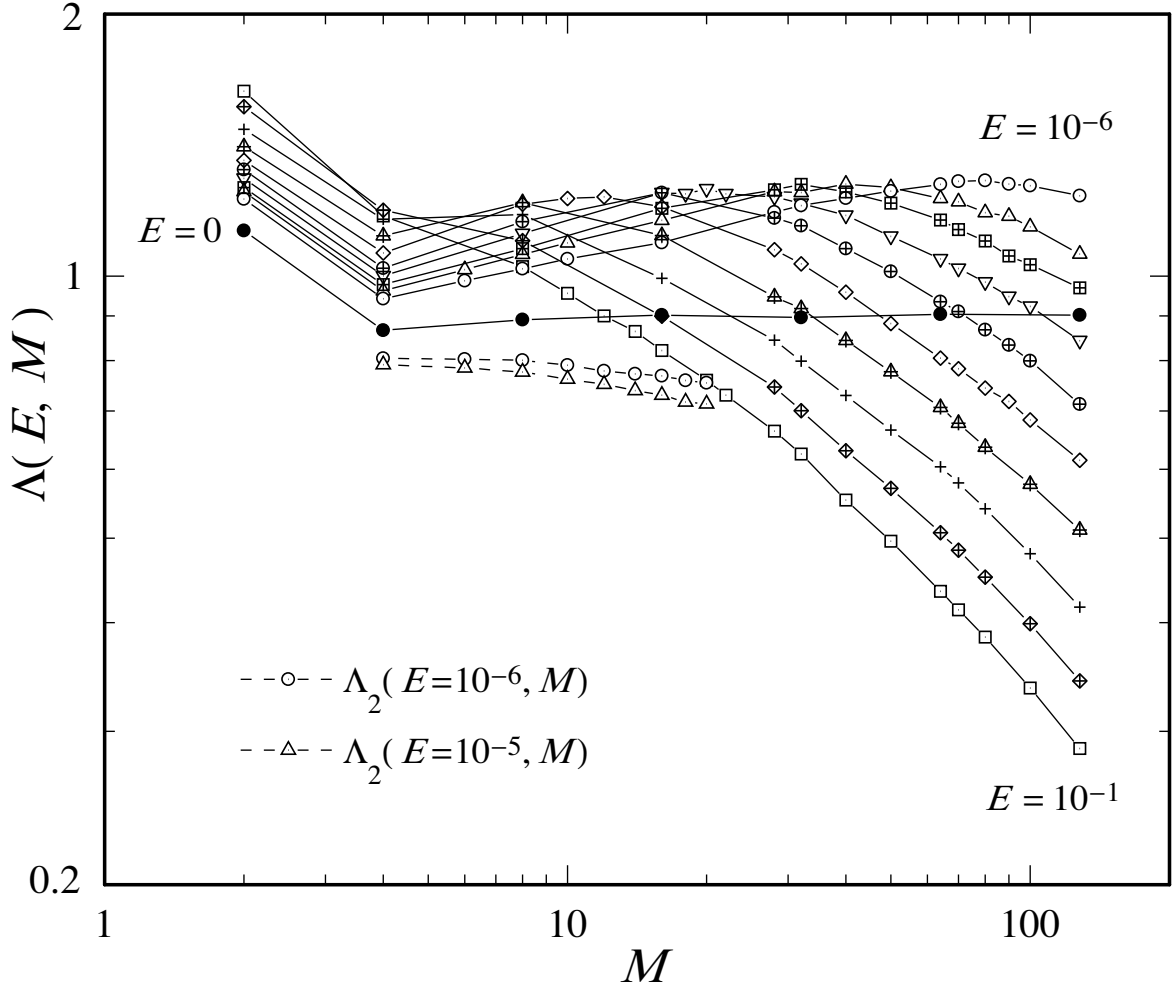}
\end{center}
\caption{The largest renormalized inverse Lyapunov exponents $\Lambda(E,M)$,
calculated for long stripes of width $M$ and various energies near the band 
center (full line). The energies range from $\log|E| = -1, -1.5, ..., -5$,
and additional $\log E = -6$. 
Dashed lines connect points of the second largest ILE, 
$\Lambda_2(E,M)$, for energies $E=10^{-5},10^{-6}$, and $M\le 20$.
} \label{ILE}
\end{figure}

In order to calculate the localization length of the infinite two dimensional
system, finite-size scaling analysis of MacKinnon and Kramer \cite{MacKinnon81}
(FSS) is applied to TMM of Pichard and Sarma \cite{Pichard81}. 
In this analysis, inverse Lyapunov exponents (ILE) of the transfer matrix 
of the ABD model \cite{Eilmes98,Brouwer99} of a long quasi one dimensional 
strip of width $M$ are calculated for several energies near the band center
and one parameter scaling analysis applied to the largest ILE, 
from which the correlation length of 2D system is calculated. 
The scaling analysis consists of assuming that the change of the 
largest ILE, $\Lambda(E,M)$, 
due to rescaling $M\to b M$ can be compensated by an appropriate change
of energy, after which $\Lambda$ will remain the same, which implies that 
\cite{MacKinnon81}
\begin{equation}\label{FSS}
  \Lambda(E,M) = \Lambda(M/\xi(E)). 
\end{equation}
Figure~\ref{ILE} shows the calculated $\Lambda(E,M)$ for several energies $E$ 
close to the band center and for various strips up to $128$ sites wide. 
The figure also gives the second largest (dashed line) renormalized ILE, 
$\Lambda_2$, for the two lowest non-zero energies studied 
($E=10^{-5}$ and $10^{-6}$) and for $M\le 20$.
All values are obtained with relative error of $1\%$ or better.

At $E=0$ and for $M$ even, all ILE become doubly degenerate 
due to the presence of chiral symmetry \cite{Miler96,Brouwer99}, 
and they scale linearly with $M$ for $M \ge 16$, reflecting 
the scale invariance of $\Lambda$ characteristic of 
a critical state.
To see the effect this degeneracy has on scaling properties of ILE,
we should recall that ILE of transfer matrices of disordered systems
repel each other in general \cite{Beenakker97} and become 
self-averaging quantities for sufficiently long stripes \cite{Pichard81}. 
The symmetry now enforces ``dimerization'' of pairs of ILE, acting as an 
effective attractive force between each pair of ILE that become degenerate
at zero energy, as in Fig.~\ref{ILE}, where, for a fixed $M$, 
$\Lambda$ and $\Lambda_{2}$ are closer together for the smaller energy.  
The largest ILE thus decreases in a strip of width $M$ on approaching the band 
center, as in models with chiral symmetry studied in Ref.~\cite{Miler96}.
On the other hand, at any given energy, every pair of ILE becomes more 
repelling with increasing of $M$.  Such increased number of ILE thus 
diminishes the attractive effect of the symmetry, and, depending on 
the relative strengths of the two effects, there are two regimes:
(i) for smaller $M$, the largest ILE increases approximately 
linearly with $M$, and, since the slope of the rise changes with energy, 
FSS cannot be done; (ii) for sufficiently large $M$, on the other hand, 
the repulsion among ILE due to disorder dominates and FSS is possible.

\begin{figure}[tb]
\begin{center}
\includegraphics*[width=6.1in]{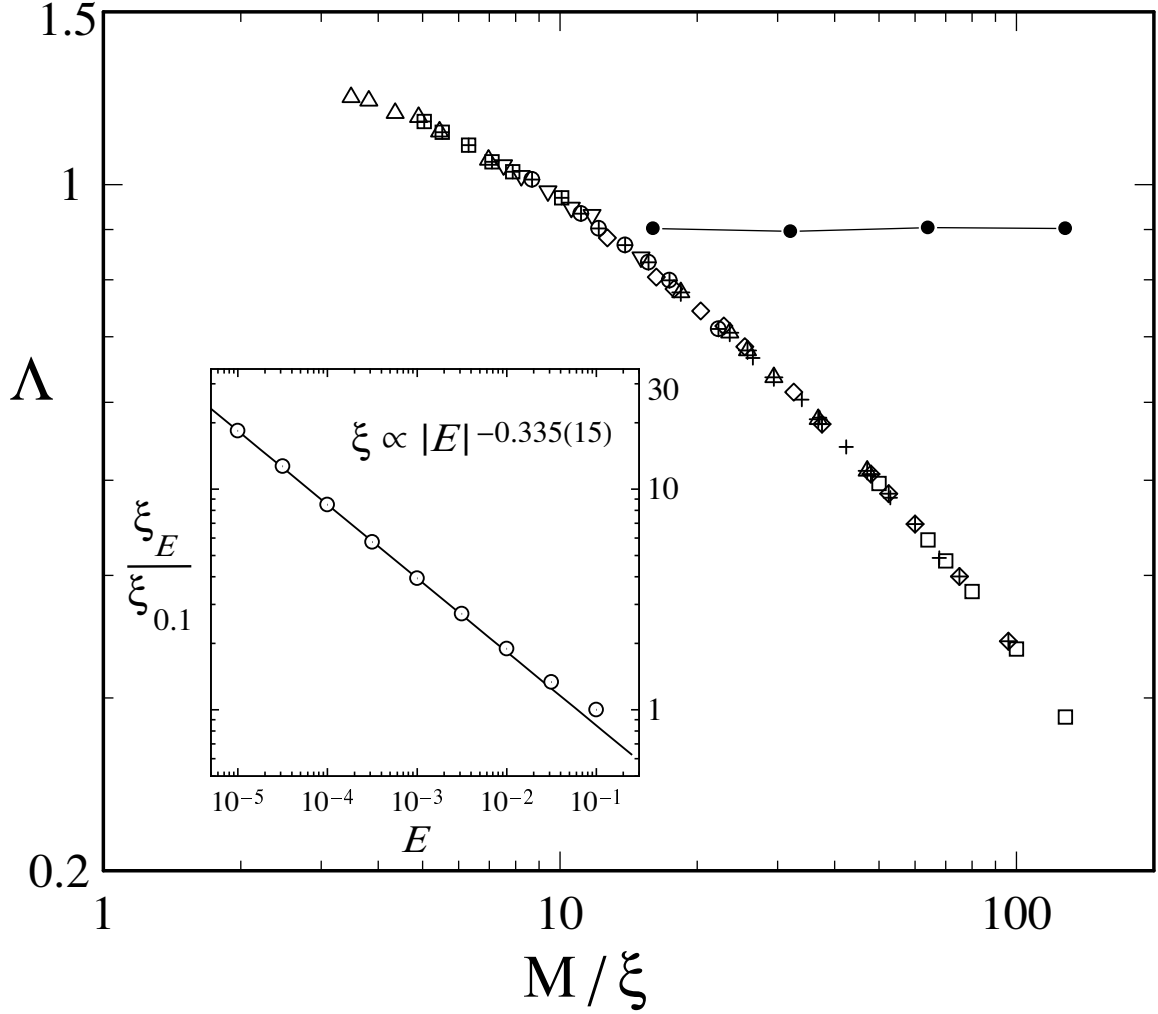}
\end{center}
\caption{One parameter universal function $\Lambda(M/\xi(E))$ for the ABD
model.  The four separate points are $\Lambda(E=0,M)$ for $M=16,32,64,128$,
corresponding to the band center critical state.
The inset shows the calculated localization length $\xi(E)$ in
units of the localization length $\xi(E=0.1)$, with error bars smaller 
than the symbol size.
} \label{universal}
\end{figure}

The one-parameter universal function $\Lambda(M/\xi(E))$, obtained for 
system sizes $M\ge 50$ for energies $|E|>10^{-4}$, and for system sizes 
$M\ge 64$ for energies $10^{-5}\le |E| \le 10^{-4}$, is presented 
in Fig.~\ref{universal}.  The obtained localization length $\xi(E)$, in units 
of the localization length of the smallest energy studied ($E=0.1$), 
is shown in the inset of the same figure.
Calculated $\xi(E)/\xi(0.1)$ is fitted to the powerlaw 
for energies $10^{-5}\le E \le 10^{-3}$ , and the result suggests powerlaw 
diverging localization length at the band center, with the exponent
\begin{equation} \label{NU}
  \nu\,=\,0.335\pm 0.034 ,
\end{equation}
in agreement with the result (\ref{NU'}) of Sec.~\ref{MULTIFRACTALITY}.

In Ref.~\cite{Eilmes98}, the authors also calculated one-parameter 
scaling curves for energies further away from the band center than in 
this work and found that there is a universal curve for a large set of 
ABD systems with different $w$ that they studied.
The obtained one-parameter function suggested that $\Lambda(x)$ either grows 
indefinitely when $E\to 0$, which is similar to the two dimensional ASD model
\cite{MacKinnon81} or becomes constant.  
If this result were correct, however, for the smaller energies than studied 
there, it would imply that localization length at the band center either 
diverges logarithmically, as in the 2D ASD model, 
or that it has a finite limiting value.
We see therefore that, although correct in the energy range 
studied in Ref.~\cite{Eilmes98}, the picture of a single scaling curve for 
different $w$ from the same work should break down for smaller energies, 
since the powerlaw diverging $\xi$ requires that 
$d\ln\Lambda / d\ln M$ goes to zero when $E\to 0$, and separate one-parameter
scaling curves exist for each $w$.

Some additional analysis of the results can be done by introducing 
$\Lambda_{max}(M)$, defined as the maximal $\Lambda(E,M)$
for a given strip width $M$.  Importance of this quantity 
comes from the fact that points where $\Lambda$ reaches its maximum 
for various energies cannot be described by FSS, and, at the same time,
$\Lambda_{max}$ limits by its definition possible values that $\Lambda(x)$
(obtained from the FSS) can have.
The main observation is that $\Lambda_{max}$ seems to grow slower than linearly 
in Fig.~\ref{ILE}.  Linear (or faster) growth would imply existence of an 
additional energy scale below which FSS would break down for all 
large $M$ and $\Lambda(E,M)$ would grow linearly and indefinitely with $M$
in the same figure, implying the existence of a whole band of extended states
where one-parameter scaling would not hold. 
Slower than linear growth of $\Lambda_{max}$ seen, therefore, furthermore 
suggests the existence of a single critical point at the band center of 
the system on an infinite square lattice.

\section{Conclusions} \label{CONCLUSIONS}

In conclusions, main results of this work are summarized.
Density of states of finite size systems are calculated and shown the validity 
of Gade's expression (\ref{Gade}).  
The calculated part of the DoS for the system on an infinite
square lattice, however, suggests a different dependence on energy near
the band center, 
\begin{equation}
  \rho(E)\,=\,C {1\over {\sqrt |E|}}\,\exp\left(-\kappa\sqrt{-\ln |E|}\right),
\end{equation}
with $\kappa=1.345\pm 0.005$ and $C=1.30\pm 0.03$.

Other calculated quantities share in common a qualitative feature of a 
discontinuous change near or at the band center.  
The nearest neighbor level spacing distributions,
for instance, between the state closest to the band center and the next one
seem to be distinctly different than the level spacing distributions between 
other neighboring states.  This was argued to be connected to the chiral
symmetry of the model, which places the two states closest to $E=0$
at the (high energy) end of the spectrum.   
These two states are furthermore found to play a crucial role in 
explaining the discontinuous change of the scaling properties of IPN 
near the band center.
Extrapolation to the limit of infinite square lattice then led to the two
different spectra of generalized dimensions $D_q$ --- one for $E\ne 0$
(present at the length scales smaller than the localization length),
and another one for $E=0$ (present at all length scales).  Finite size
scaling associated with the effects that finite $L$ has on the critical
band center states of the infinite system led to the value 
$\nu = 0.35\pm 0.01$ of the critical exponent of the localization length.

Additional scaling analysis of ILE of transfer matrices of long quasi-1D
systems gave the value $\nu\,=\,0.335\pm 0.034$.
This suggests that there is only one critical state at the band center 
with all other states localized in the system on the infinite square lattice,
in agreement with findings of Ref.\ \cite{Soukoulis84,Eilmes98}. 
One parameter universal function $\Lambda(x)$ is calculated and 
another discontinuity found, since $\Lambda(x(E\to0)) \ne \Lambda(E=0,M)$.

A puzzling feature of the critical exponent $\nu$ of the ABD
as well as of the RDF model \cite{Hatsugai97,Morita97} is their apparent 
disagreement with the rigorous theorem of 
Chayes and Chayes {\it et.\ al} \cite{ChayesandChayes86}, 
which states that $\nu\ge 1$ in two-dimensional quantum disordered systems 
in general.  This question, however, requires further study and will be
addressed elsewhere.

\acknowledgements

This work is partially supported by the Department of Physics and Astronomy
at Michigan State University.  Author is thankful to S.\ D.\ Mahanti for
suggestions on improving the manuscript and to S.\ A.\ Trugman, R.\ Bhatt,
T.\ A.\ Kaplan, V.\ Zelevinski, M.\ Mostovoy, I.\ Herbut, D.\ Mulhall,
B.\ Nikoli\'c and M.\ Milenkovi\'c for inspiring discussions.

\break

\break 

\break

\break

\break

\break

\break

\break

\break 

\break

\begin{thebibliography}{99}
\bibitem{Inuifootnote}
It is assumed here, without loss of generality, that $N_A \ge N_B$.  
The $m$ zero-energy eigenstates still remain if arbitrary hoppings 
are introduced among sites of the sublattice $B$ \cite{Inui94}.
\bibitem{preliminary}
Some preliminary results indicate that $D_1(s)$ has characteristics of the
critical distribution, while other $D_i(s)$ have exponential tail 
characteristic for localized states.
\end{thebibliography}
\end{document}